\documentclass[a4paper,10pt,twoside]{cpcstyle}

\usepackage{multicol}
\usepackage{graphicx}
\usepackage{booktabs}
\usepackage{amssymb,bm,mathrsfs,bbm,amscd}
\usepackage[tbtags]{amsmath}
\usepackage{lastpage}

\def\jpsi{$J/\psi$ }
\def\psiprime{$\psi'$ }
\def\t0{$t_0$ }

\begin{document}

\fancyhead[co]{\footnotesize N.~Berger~ et al: Trigger efficiencies at BES III}


\title{Trigger efficiencies at BES III}

\author{%
      N. Berger$^{1;1)}$\email{nberger@mail.ihep.ac.cn}
      \quad K. Zhu$^{1;2)}$\email{zhuk@mail.ihep.ac.cn}
\quad Z.A. Liu$^{1}$%
\quad D.P. Jin$^{1}$%
\quad H. Xu$^{1}$%
\quad W.X. Gong$^{1}$%
\quad K. Wang$^{1}$%
\quad G. F. Cao$^{1}$%
}
\maketitle

\address{%
$^1$: Institute of High Energy Physics, Chinese Academy of Sciences, Beijing 100049, China\\
}

\begin{abstract}
Trigger efficiencies at BES III were determined for both the \jpsi and \psiprime
data taking of 2009. Both dedicated runs and physics datasets are used; efficiencies are
presented for Bhabha-scattering events, generic hadronic decay events involving charged 
tracks, dimuon events and $\psi' \longrightarrow \pi^+\pi^-J/\psi$, $J/\psi \longrightarrow
\ell^+\ell^-$ events. The efficiencies are found to lie well above 99\% for all relevant
physics cases, thus fulfilling the BES III design specifications. 
\end{abstract}



\begin{multicols}{2}

\section{Introduction}

The Beijing Spectrometer III (BES III) experiment at the  Beijing Electron-Positron Collider II (BEPC II) was designed to make use of the world's largest luminosity in the $\tau$-charm energy region for studies of the light hadron spectrum, charmonia, open charm and $\tau$-lepton production. In order to exploit the luminosities of up to $10^{33}~\textrm{cm}^{-2}\textrm{s}^{-1}$, BES III requires a fast data acquisition as well as an efficient and highly selective trigger system.

The BES III trigger system as used for the 2009 data taking combines information from the electromagnetic calorimeter (EMC), the main
drift chamber (MDC) and the time-of-flight system (ToF) to select the interactions of interest for readout\footnote{The system also incorporates trigger conditions based on the muon counter (MUC) and coincidences between subsystems (\emph{match}-trigger) 
these were however not used in the 2009 running.}.
A detailed description of the system can be found in \cite{BES3Det, Liu}.

The subsystems provide a set of \emph{trigger conditions} each (48 in total, see 
table \ref{tab1}), which are combined to up to 13 \emph{trigger channels} 
by the global trigger logic (GTL). In addition, there is a \emph{random trigger} and two different \emph{prescales}\footnote{Not used for the 2009 data taking.}. If any enabled trigger channel is active, the event will be read out.
A typical \emph{trigger menu} (combination of trigger conditions to trigger channels) used in physics data
taking is shown in Table \ref{tab2}.

There are two groups of EMC based trigger conditions, one is based on clusters (localized deposits of energy above a threshold) and their distribution in polar and azimuthal angle. The second group is based on the total energy deposited in the calorimeter. ToF conditions are based on the number and relative location in azimuth of signals in the scintillator bars. MDC conditions finally are derived from the patterns of active cells in the drift chambers. Hits in four neighbouring layers within a \emph{super layer} are flagged as \emph{track segments} if they are compatible with the trajectory of a particle originating from the beam line. This matching is performed in four super layers. If there are matching segments from the innermost three super layers, this is considered a \emph{short track}, if a segment is also present in the fourth super layer (which restricts the track to be contained in the barrel region), a \emph{long track} is formed.

Events not seen by the trigger are lost forever; thus a high efficiency is desirable. On the other hand,
the capacity of the readout system is limited; the trigger should also be highly selective. The trigger
efficiency needs to be considered in the overall normalisation of any physics analysis\footnote{A detailed simulation of the BESIII trigger for use with Monte Carlo simulation events is currently being implemented, has however not reached full maturity at the time of writing.}. For the precursor experiments BES I and II, trigger efficiencies were determined both using data \cite{Yu1995, Huang2001} and simulations \cite{Liu2001, Shi2004}. In the present paper, trigger efficiencies are determined from data. The methods employed are introduced in the next section. This is followed by a description of the data sample and selection employed. The results of the study in Section 4 are followed by a brief conclusion.

\section{Efficiency determination}

As all the data available for trigger studies have actually been triggered, the main challenge in the efficiency determination is to reduce any bias caused by this to a minimum. Ideally, the trigger efficiency would be determined from random trigger events, this however suffers from very low statistics as soon as a physics selection is applied. Instead, samples triggered by independent trigger channels are used. The MDC and ToF based trigger conditions are checked using events triggered by trigger Channel 11, which is based on EMC conditions only. Conversely, EMC based trigger conditions are checked using the MDC and ToF condition based Channels 2 and 3\footnote{For the \jpsi running, Channel 3 was not active, so only Channel 2 was used.}. In addition to the standard \psiprime and \jpsi runs, special runs with a trigger set-up geared especially at efficiency determination were taken in May and July of 2009 respectively. They will be designated as \emph{trigger test runs} in the following in order to discern them from standard data taking. The experience gained during the \psiprime running allowed to set up such monitor triggers also for the endcap conditions in the \jpsi test run.


\subsection{Trigger condition efficiencies}

As all the trigger conditions used in the 2009 running are determined from a single sub-detector, a data sample triggered by a trigger channel not using that sub-detector can be used as a reference:
\begin{equation}
	\epsilon = \frac{N_{(sel, channel, condition)}}{N_{(sel, channel)}}.
\end{equation}
Here $N$ stands for the number of events, the label $sel$ for events passing the physics selection, $channel$ for events having triggered the reference channel and $condition$ for events with the trigger condition under study active.

\subsection{Timing dependence}

The high bunch crossing frequency of 125~MHz at BEPC II is not resolved by the trigger electronics. This, together with the different latencies, time resolutions and latch times of the trigger subsystems poses a challenge to trigger
timing. Checks of the trigger timing and the proper coincidence of subsystem signals are thus of utmost importance.

The BES III trigger system is not able to resolve individual bunch crossings. Data in a broad timing window around the time of the trigger are read out and the exact event timing (\t0) is determined off-line. It is thus very important that both the position of the readout window with regard to the trigger signal is correct as well as that the trigger conditions actually coincide (as opposed to being active somewhere in the readout window).

\begin{center}
\tabcaption{ \label{tab1}  Trigger conditions. Note that $N$ used in conditions 39 and 43 was set to 16 for the 2009 running. Trigger conditions 23 to 37 are reserved for the muon system trigger and a trigger matching information from different subdetectors.}
\footnotesize
		\begin{tabular*}{80mm}{rll}
		\toprule
		 & Short Name & Condition \\
		\hline 
		\multicolumn{3}{c}{Electromagnetic calorimeter (EMC)} \\
		\hline
		0  & \verb|NClus.GE.1| & Number of clusters $\geq 1$ \\
		1  & \verb|NClus.GE.2| & Number of clusters $\geq 2$ \\
		2	 & \verb|BClus_BB|   & Barrel cluster back-to-back \\
		3  & \verb|EClus_BB|   & Endcap cluster back-to-back \\
		4  & \verb|Clus_Z|     & Cluster balance in $z$ direction \\
		5  & \verb|BClus_Phi|  & Barrel cluster balance in $\varphi$\\
		6  & \verb|EClus_Phi|  & Endcap cluster balance in $\varphi$\\
		7  & \verb|BEtot_H|    & Barrel energy, high threshold \\
		8  & \verb|EEtot_H|    & Endcap energy, high threshold \\
		9  & \verb|Etot_L|     & Total energy, low threshold \\
		10 & \verb|Etot_M|     & Total energy, medium threshold \\
		11 & \verb|BL_EnZ|     & Energy balance in $z$ direction \\
		12 & \verb|NBClus.GE.1|& Number of barrel clusters $\geq 1$ \\
		13 & \verb|NEClus.GE.1|& Number of endcap clusters $\geq 1$\\
		14 & \verb|BL_BBLK|    & Barrel energy block balance \\
		15 & \verb|BL_EBLK|    & Endcap energy block balance \\
		\hline 
		\multicolumn{3}{c}{Time of flight system (ToF)} \\
		\hline
		16 & \verb|ETOF_BB|    & Endcap ToF back-to-back \\
		17 & \verb|BTOF_BB|    & Barrel ToF back-to-back \\
		18 & \verb|NETOF.GE.2| & Number of endcap ToF hits $\geq 2$\\
		19 & \verb|NETOF.GE.1| & Number of endcap ToF hits $\geq 1$\\
		20 & \verb|NBTOF.GE.2| & Number of barrel ToF hits $\geq 2$\\
		21 & \verb|NBTOF.GE.1| & Number of barrel ToF hits $\geq 1$\\
		22 & \verb|NTOF.GE.1|  & Number of ToF hits $\geq 1$\\
		\hline
		\multicolumn{3}{c}{Main drift chamber (MDC)} \\
		\hline
		38 & \verb|STrk_BB|    & Short tracks back-to-back \\
		39 & \verb|STrk.GE.N|  & Number of short tracks $\geq N$ \\
		40 & \verb|STrk.GE.2|  & Number of short tracks $\geq 2$ \\
		41 & \verb|STrk.GE.1|  & Number of short tracks $\geq 1$ \\
		42 & \verb|LTrk_BB|    & Long tracks back-to-back \\
		43 & \verb|LTrk.GE.N|  & Number of long tracks $\geq N$ \\
		44 & \verb|LTrk.GE.2|  & Number of long tracks $\geq 2$ \\
		45 & \verb|LTrk.GE.1|  & Number of long tracks $\geq 1$ \\	
		46 & \verb|ITrk.GE.2|  & Number of inner tracks $\geq 2$ \\
		47 & \verb|ITrk.GE.1|  & Number of inner tracks $\geq 1$ \\	
		\bottomrule
\end{tabular*}
\end{center}

The efficiency of EMC and MDC signals was determined relative to the ToF signals, which provide the greatest timing accuracy (minimum signal width of 4 cycles of the 40~MHz clock in the trigger, corresponding to 12 bunch crossings). The ToF reference signal (usually \verb|NBTOF.GE.1|) is required to be on in the four nominal cycles with regard to the event \t0. If the trigger condition under study has its signal on in any of these four cycles, this is treated as a coincidence and counted in the numerator of the efficiency fraction. In order to minimize the effects of the limited readout window size, the event \t0 is also required to lie within the nominal range. 

\subsection{Trigger channel efficiencies}
\label{sec:DeterminationOfTriggerChannelEfficiencies}

The efficiency of trigger channels can be determined analogously to the efficiency of the trigger conditions if a fully independent trigger channel exists (i.e.~Channels 2 and 3, which do not depend on the EMC and Channel 11, which depends solely on the EMC). Otherwise, correlations between the channels have to be accounted for. The overall trigger efficiency is then obtained by combining all the channels, where again correlations need to be considered. 

Assuming that the readout window is wide enough, the trigger channel efficiency determined for the independent channels from the trigger decision readout is trigger timing independent. This is not the case for the mathematical combination; checks for Bhabha and Hadron data however indicate that the timing effects are small, certainly below 0.5\% for the studied event topologies.

\section{Sample selection}

The selections employed in the trigger studies are less refined than those employed in physics analyses. The intention is not so much the selection of a very pure sample but the identification of a large number of events that share the main features (as far as the trigger is concerned) with the signal events.

In general, objects with a polar angle $\vartheta$ in the range $|\cos \vartheta| < 0.97$ are considered. The barrel region is defined as $|\cos \vartheta| < 0.8$, the endcap region as $0.83 < |\cos \vartheta| < 0.97$. Tracks in the main drift chamber are required to have a distance of closest approach to the run-averaged interaction point of less than 1~cm in the radial direction and less than 5~cm along the beam direction to be considered in the selection.

\subsection{Bhabha selection}

The Bhabha selection requires two calorimeter clusters with an energy within 10\% of the beam energy and an opening angle larger than $166^\circ$.
In addition, two charged tracks with an opening angle of more than $175^\circ$ are required.

\subsection{Charged hadron selection}
In the inclusive charged hadron selection, two or more charged tracks are required to be present in the event. If there are exactly two tracks, their opening angle is required to be less than $170^\circ$ in order to suppress Bhabha and dimuon events.

\begin{center}
	\tabcaption[Trigger settings for the \psiprime running]{\label{tab2}Trigger settings for the 2009 \psiprime running. Channel 0 is designed for
	endcap Bhabha events, Channels 1 to 5 for events with charged particles in the barrel region and Channel 11 for all-neutral events. For the \jpsi runs, the same set-up without Channel 3 was used.}
	\footnotesize
		\begin{tabular*}{80mm}{rl}
			\toprule
			Channel & Conditions \\
			\hline
			0 & \verb|STrk_BB && NETOF.GE.1 && NEClus.GE.1|\\
			1 & \verb|NLTrk.GE.2 && NBTOF.GE.2 && NBClus.GE.1|\\
			2 & \verb|NLTrk.GE.2 && NBTOF.GE.2|\\
			3 & \verb|LTrk_BB && NBTOF.GE1|\\
			4 & \verb|NLTrk.GE.1 && NBTOF.GE.1 && Etot_L|\\
			5 & \verb|NLtrk.GE.2 && NBTOF.GE.1 && NBClus.GE.1|\\
			9& Random trigger at 60~Hz\\
			11& \verb|NClus.GE.2 && Etot_M|\\
			\bottomrule
		\end{tabular*}
\end{center}

\subsection{Dimuon selection}
The dimuon selection requires two good charged tracks with opposite
charge and a spacial angle between them larger than
$177.6^\circ$. In addition, we require that the momentum of each
track is less than 2~GeV/$c$ and the deposited energy in the EMC is less than
0.7~GeV. The total four-momentum $(E/c,p_x,p_y,p_z)$ is required to fall in the
$(2.8\sim 3.3, -0.1\sim 0.1, -0.1\sim 0.1, -0.2\sim 0.2)$~GeV/$c$ region for \jpsi data (for the \psiprime data, the energy region is $3.2\sim4$~GeV/$c$) after assuming each track is a muon.

\subsection{$\psi' \longrightarrow \pi^+\pi^-J/\psi$ Selection}
The $\psi' \to \pi^+\pi^-J/\psi$ selection requires four
good charged tracks, two positive and two negative. The good charged
tracks are defined as in the previous selections. Then we assume the softer
tracks of each charge are pions and harder tracks are leptons. After that
assumption we also require the recoil mass of the two pions to fall into the region
$(3.08,3.12)$~GeV/$c^2$, the invariant mass of two leptons is required to be in the region
$(3.0,3.2)$~GeV/$c^2$, and the total energy of the event should be in the range 
$(3,4)$~GeV.

\section{Results}
\label{sec:Results}

\subsection{Trigger condition efficiencies}

Using the methods described in the previous sections, the efficiency for the trigger conditions used in the trigger menus was determined. The numerical results are shown in Table \ref{tab3} and \ref{tab4}. In Figure \ref{fig1} the threshold behaviour of various EMC conditions is shown.
Figure \ref{fig2} depicts the dependence of the efficiency of an MDC condition (two or more long tracks) on the number of off-line reconstructed tracks. Figure \ref{fig3} does the same for a ToF condition (two or more hits in the barrel). The figures were derived from the \jpsi trigger test run data using the inclusive charged hadron selection, they are, however, virtually indistinguishable from the corresponding figures derived from any of the other data sets.

\begin{center}
\includegraphics[width=6.0cm]{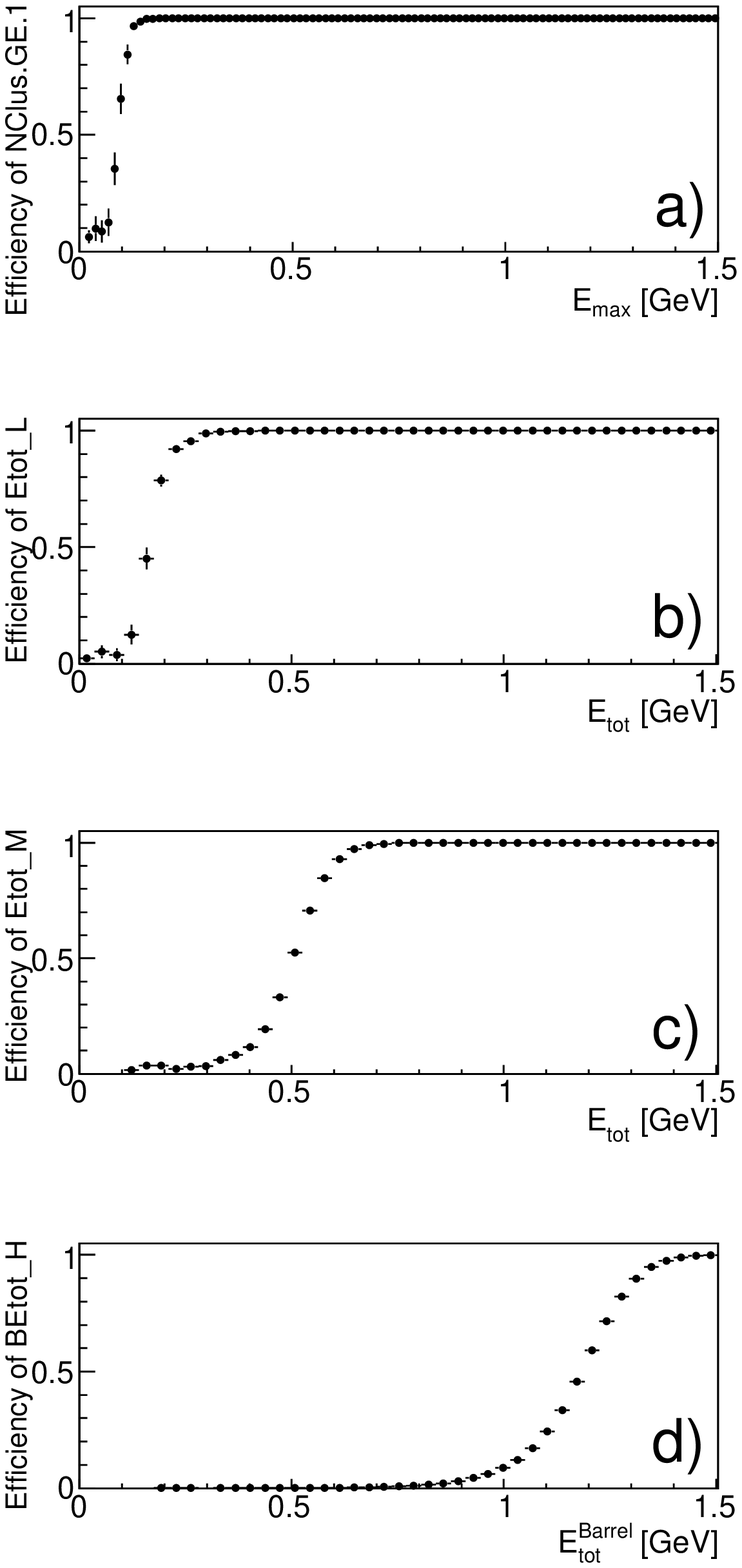}
\figcaption{\label{fig1} Threshold behaviour of the EMC trigger conditions in the \jpsi running. From top to bottom: a) Efficiency of the NClus.GE.1 condition versus maximum cluster energy, b) Efficiency of the low energy threshold versus total EMC energy, c) Efficiency of the medium energy threshold versus total EMC energy, d) Efficiency of the barrel high energy threshold versus barrel EMC energy.}
\end{center}

\begin{center}
\includegraphics[width=7cm]{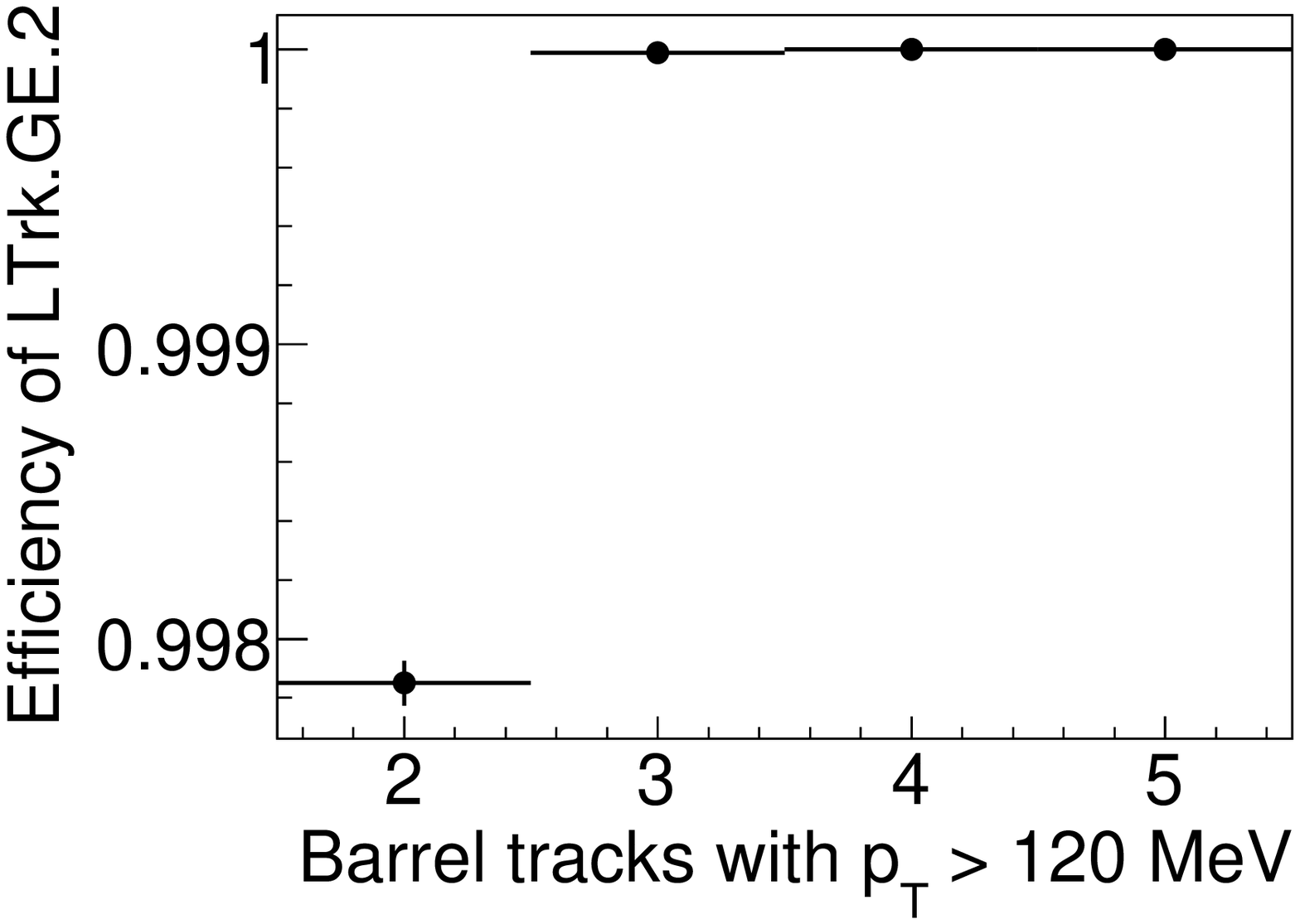}
\figcaption{\label{fig2} Efficiency of the \emph{two or more long tracks} MDC trigger condition versus the number of tracks with $p_T$ larger than 120~MeV in the barrel.}
\end{center}

\begin{center}
\includegraphics[width=7cm]{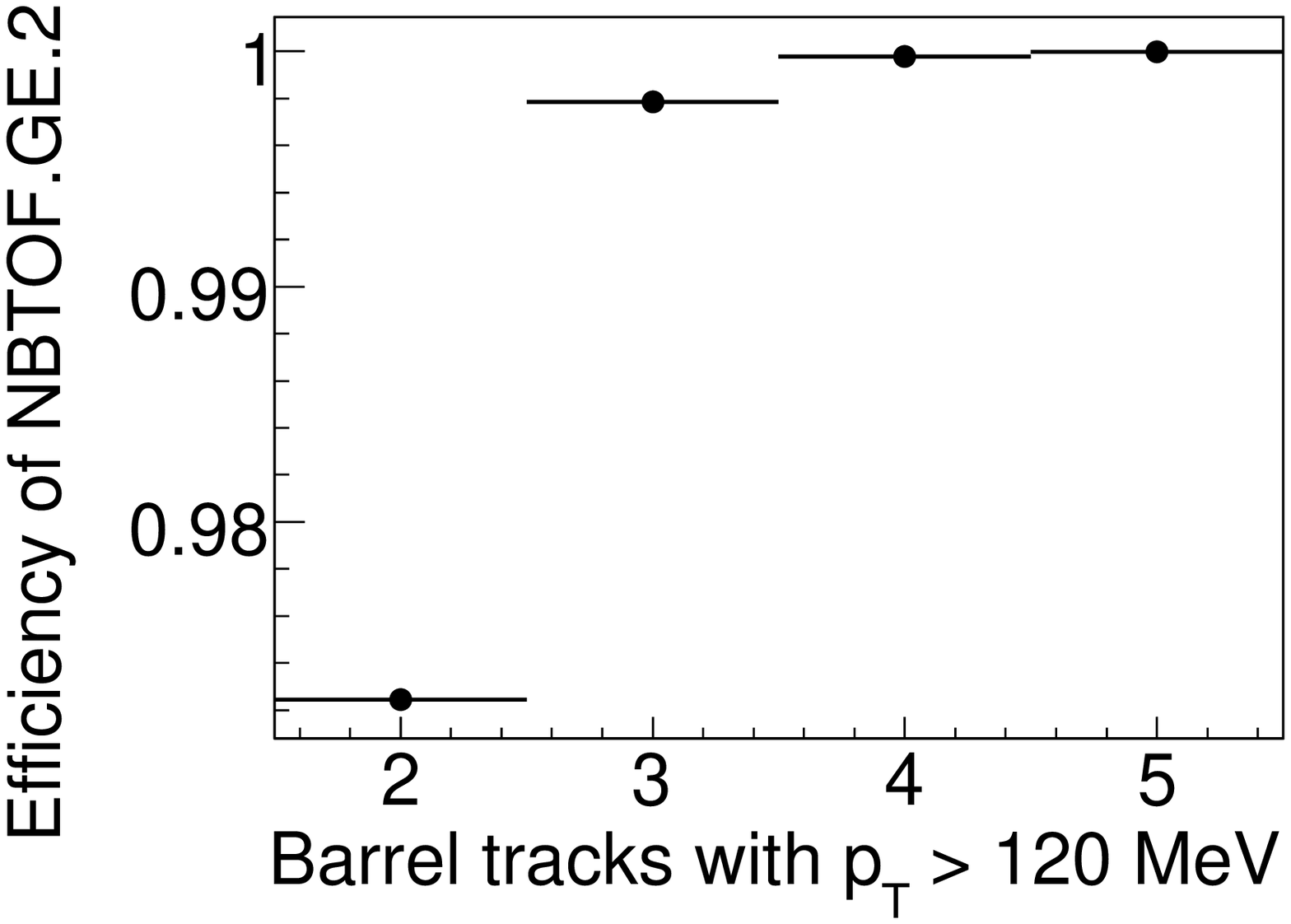}
\figcaption{\label{fig3} Efficiency of the \emph{two or more hits} ToF trigger condition versus the number of tracks with $p_T$ larger than 120~MeV in the barrel.}
\end{center}

\begin{center}
\includegraphics[width=7cm]{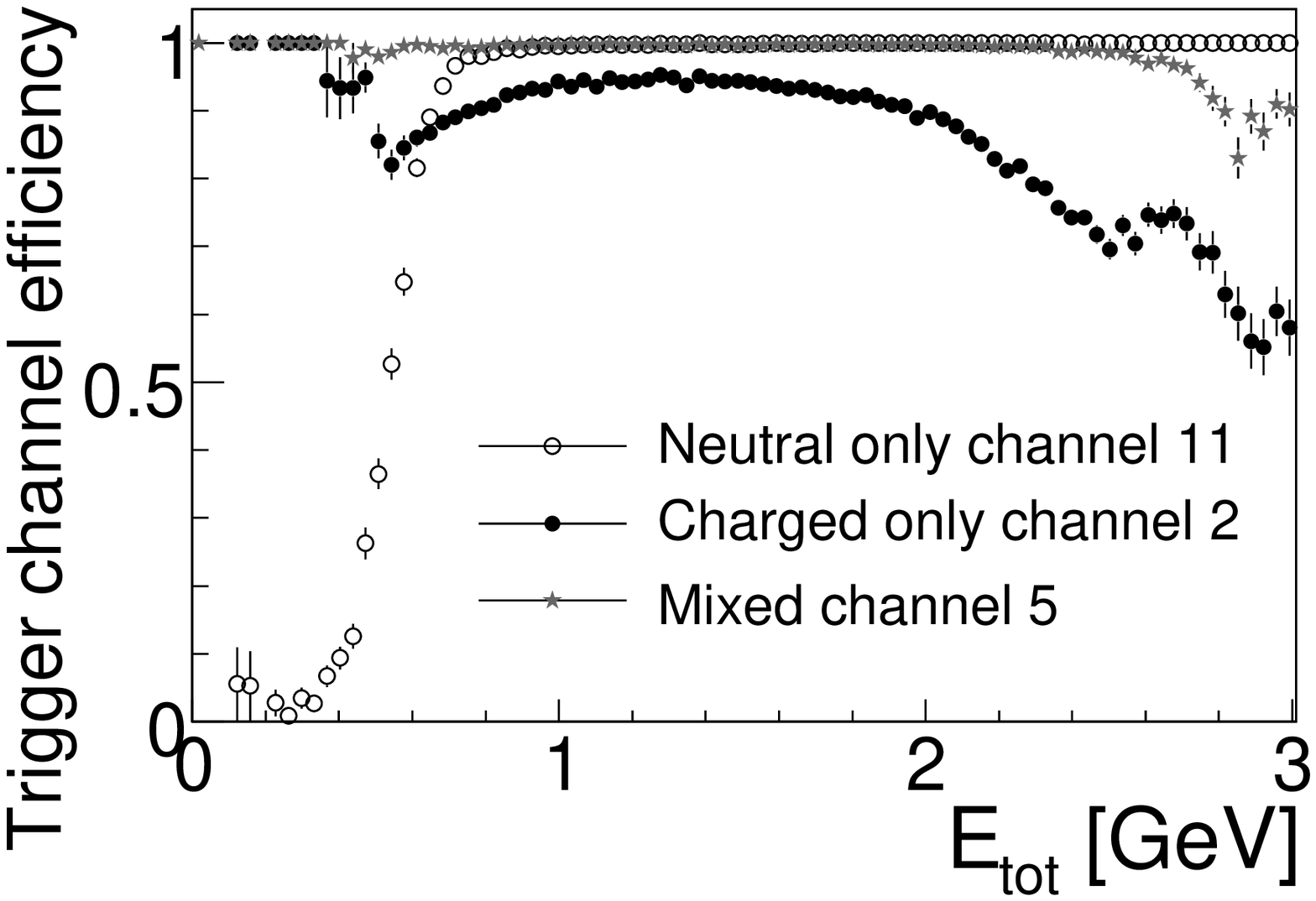}
\figcaption{\label{fig4} Efficiency of a selection of trigger channels in \jpsi running versus the total energy deposit in the EMC.}
\end{center}

\subsection{Global trigger efficiencies}

Efficiencies of the various trigger channels employed in the \psiprime and \jpsi physics run can be found in Table \ref{tab5} and \ref{tab6} respectively. Figure \ref{fig4} shows the dependence of the trigger efficiency of the all-neutral, MDC and ToF only and the minimal MDC, ToF and EMC requirement trigger channels versus the total energy deposit in the EMC. 

\subsection{Noise levels}

In order to determine noise levels and check the selectivity of the (barrel) trigger conditions, \emph{endcap} Bhabha events triggered by endcap-only trigger conditions were selected. In this sample, the fraction of events with a specific \emph{barrel} condition fulfilled is determined. This results in an upper limit of the noise ratio, as some of the conditions may also be activated by signal leaking from the endcaps into the barrel. Table \ref{tabN} shows the noise fractions in the \jpsi trigger test run.

\section{Conclusion}

The efficiency of the BES III trigger for various physics channels has been determined. For most cases, the efficiency of the full trigger menu approaches 100\% and can thus be safely neglected in physics analyses. Only for events with both a low track multiplicity and a small calorimetric energy deposit (e.g.~dimuon events), there are sizeable inefficiencies that have to be taken into account.


\end{multicols}
\begin{center}
\tabcaption{ \label{tab3}  Trigger condition efficiencies for \psiprime data taking. The efficiency for the  Bhabha events was determined taking into account the trigger signal timing; as the ToF signals serve as a reference in this method, their efficiency cannot be determined. The numbers have statistical errors of the order of 0.1\% in efficiency.}
\footnotesize
\begin{tabular*}{170mm}{@{\extracolsep{\fill}}lrrrr}
\toprule 
		Condition & Efficiency    & Efficiency & Efficiency & Efficiency \\
							&	Barrel        & Hadron     & Dimuon     & \psiprime $\rightarrow \pi^+\pi^- J/\psi$, \\
							& Bhabha        &            &            & \jpsi $\rightarrow \ell^+\ell^-$\\
							\hline
\texttt{NClus.GE.1} & 99.9 \%  & 99.2 \%  & 99.4 \%  & 99.7 \%  \\
\texttt{NClus.GE.2} & 98.8 \%  & 97.7 \%  & 96.3 \%  & 97.6 \%  \\
\texttt{Etot\_L}    &100.0 \%  & 99.0 \%  & 99.9 \%  & 99.6 \%  \\
\texttt{Etot\_M}    &100.0 \%  & 87.3 \%  & 1.4 \%  & 21.6 \%  \\
\texttt{NLTrk.GE.1} & 99.6 \%  & 99.9 \%  & 99.0 \%  & 99.8 \%  \\
\texttt{NLTrk.GE.2} & 99.7 \%  & 99.0 \%  & 98.9 \%  & 99.5 \%  \\
\texttt{NBTOF.GE.1} &          & 99.6 \%  & 99.6 \%  & 99.7 \%  \\
\texttt{NBTOF.GE.2} &          & 95.7 \%  & 96.0 \%  & 97.2 \%  \\
\bottomrule
\end{tabular*}
\end{center}
\begin{center}
\tabcaption{ \label{tab4}  Trigger condition efficiencies for \jpsi data taking. The numbers have statistical errors of the order of 0.1\% in efficiency. The number of prongs for hadronic events refers to the number of charged tracks in the barrel. For the hadronic events, the effects of timing are considered; as the ToF system serves as a reference, no ToF efficiencies were determined. The short track conditions were found to be slightly out of time for barrel events.}
\footnotesize
\begin{tabular*}{170mm}{@{\extracolsep{\fill}}lrrrrrr}
\toprule 
	Condition & Efficiency    & Efficiency 	& Efficiency & Efficiency & Efficiency & Efficiency \\
							&	Barrel        & Endcap      & 2-prong    & 4-prong    & Barrel     & Endcap     \\
							& Bhabha        & Bhabha      & Hadrons    & Hadrons    & Dimuon     & Dimuon     \\
							\hline
\texttt{NClus.GE.1} &100.0 \%  &100.0 \%  &  99.9\% &  100.0\%& 100.0 \%  & 86.3 \%  \\
\texttt{NClus.GE.2} & 99.0 \%  & 96.5 \%  &  94.3\% &  97.6\% & 96.8 \%  & 50.5 \%  \\
\texttt{Etot\_L}    &100.0 \%  &100.0 \%  &  100.0\%&  100\% &100.0 \%  & 97.9 \%  \\
\texttt{Etot\_M}    &100.0 \%  &100.0 \%  &  96.2\% &  99.1\% &  7.6 \%  &  5.0 \%  \\
\texttt{BEtot\_H}   &100.0 \%  &  0.0 \%  &  61.9\% &  50.2\% &  0.3 \%  &  0.1 \% \\
\texttt{EEtot\_H}   &  0.1 \%  &100.0 \%  &  0.3\%  &   0.1\% &  0.1 \%  &  2.0 \% \\
\texttt{NLTrk.GE.1} &100.0 \%  & 17.2 \%  &  99.8\% &  99.9\% &100.0 \%  & 32.9 \%  \\
\texttt{NLTrk.GE.2} &100.0 \%  &  0.7 \%  &  98.7\% &  99.9\% &100.0 \%  & 22.5 \%  \\
\texttt{NSTrk.GE.1} &100.0 \%  &100.0 \%  &  98.7\% &  99.0\% &100.0 \%  &100.0 \%  \\
\texttt{NSTrk.GE.2} &100.0 \%  &100.0 \%  &  98.5\% &  99.2\% &100.0 \%  &100.0 \%  \\
\texttt{NBTOF.GE.1} &100.0 \%  & 17.0 \%  &         &         &100.0 \%  & 17.2 \%  \\
\texttt{NBTOF.GE.2} & 98.2 \%  &  0.8 \%  &         &         & 97.6 \%  &  0.9 \%  \\
\texttt{NETOF.GE.1} & 22.5 \%  & 99.8 \%  &         &         & 21.8 \%  & 95.2 \%  \\
\texttt{NETOF.GE.2} &  1.6 \%  & 94.3 \%  &         &         &  1.6 \%  & 76.7 \%  \\\bottomrule
\end{tabular*}%
\end{center}
\begin{multicols}{2}

\end{multicols}
\begin{center}
	\tabcaption{\label{tab5}Global trigger efficiencies for \psiprime running. The statistical errors are below 0.1\% in efficiency.}
	\footnotesize
		\begin{tabular*}{170mm}{@{\extracolsep{\fill}}lrrrr}
			\toprule
			Trigger    & Barrel     & Hadron     & Dimuon      & \psiprime $\rightarrow \pi^+\pi^-J/\psi$, \\
			channel    & Bhabha     & efficiency & efficiency  &	\jpsi $\rightarrow \ell^+ \ell^-$       \\
			group      & efficiency &            &             &	efficiency \\
		\hline 
		Channel 0  & 0.3 \% & 31.8 \% & 0.4 \%  & 13.7 \% \\
		Channel 1  & 93.5\% & 92.1 \% & 95.0 \% & 94.7 \% \\
		Channel 2  & 93.5\% & 93.1 \% & 98.5 \% & 96.7 \% \\
		Channel 3  & 98.9\% & 72.8 \% & 98.6 \% & 93.5 \% \\
		Channel 4  & 99.0\% & 98.7 \% & 98.7 \% & 99.2 \% \\
		Channel 5  & 98.9\% & 96.1 \% & 98.0 \% & 97.2 \% \\
		Channel 11 & 98.5 \% & 85.9 \% &  1.3 \% & 21.1 \% \\
		\hline
		Barrel charged & 99.0 \% & 99.7 \% & 98.7 \% & 99.6 \% \\
		Endcap charged &  0.3 \% & 31.8 \% & 0.4 \% & 13.7 \% \\
		Neutral        & 98.5 \% & 85.9 \% & 1.3 \% & 21.1 \% \\
		\hline
		Total          &99.99 \% & 99.97 \% & 98.7 \% & 99.7 \% \\
		\bottomrule
		\end{tabular*}
\end{center}
\begin{center}
	\tabcaption{\label{tab6}Global trigger efficiencies for \jpsi running. The statistical errors are below 0.1\% in efficiency.}
	\footnotesize

		\begin{tabular*}{170mm}{@{\extracolsep{\fill}}lrrrrr}
	\toprule
			Trigger    & Barrel     & Endcap    & Hadron     & Barrel      & Endcap \\
			channel    & Bhabha     & Bhabha 		& efficiency & Dimuon      &	Dimuon  \\
			group      & efficiency & efficiency&            &	efficiency  &	efficiency\\
		\hline
		Channel 0      &  0.14 \% & 99.8 \% & 25.6 \% &  0.2 \% & 94.3 \% \\
		Channel 1			 & 98.2 \%  &  0.0 \% & 98.5 \% & 97.6 \% &  0.0 \% \\
		Channel 2      & 98.2 \%  &  0.1 \% & 98.6 \% & 97.6 \% &  0.2 \% \\
		Channel 4      & 99.96 \% &  0.7 \% & 99.98 \%& 99.9 \% &  3.3 \% \\
		Channel 5      & 99.99 \% &  2.9 \% & 99.85 \%& 99.9 \% &  5.5 \% \\
		Channel 11     & 99.0 \%  & 96.5 \% & 97.7 \% &  7.3 \% &  2.5 \% \\
		\hline
		Barrel charged & 99.99 \% &  2.9 \% & 99.99 \%& 99.9 \% & 5.6 \% \\
		Endcap charged &  0.14 \% & 99.8 \% & 25.6 \% & 0.2 \% & 94.3 \% \\
		Neutral        &  99.0 \% & 96.5 \% & 97.7 \% & 7.3 \% & 2.5 \% \\
		\hline
		Total          &100.0 \% & 99.99 \% & 100.0\% & 99.94 \% & 94.8 \% \\
	\bottomrule
		\end{tabular*}
\end{center}
\begin{multicols}{2}

\begin{center}
\tabcaption{ \label{tabN}  Barrel trigger condition noise levels as determined from endcap Bhabha events in the \jpsi trigger test run.}
\footnotesize
		\begin{tabular*}{80mm}{@{\extracolsep{\fill}}lrl}
		\toprule
		 & Short Name & Noise fraction \\
		\hline 
		\multicolumn{3}{c}{Electromagnetic calorimeter (EMC)} \\
		\hline
		2	 & \verb|BClus_BB|   & \phantom{1}0.0014\% \\
		5  & \verb|BClus_Phi|  & \phantom{1}0.011\%\\
		7  & \verb|BEtot_H|    & \phantom{1}0.017\% \\
		12 & \verb|NBClus.GE.1|& \phantom{1}0.19\% \\
		14 & \verb|BL_BBLK|    & \phantom{1}0.036\% \\
		\hline 
		\multicolumn{3}{c}{Time of flight system (ToF)} \\
		\hline
		17 & \verb|BTOF_BB|    & \phantom{1}0.099\% \\
		20 & \verb|NBTOF.GE.2| & \phantom{1}0.80\%\\
		21 & \verb|NBTOF.GE.1| & 17.1\%\\
		\hline
		\multicolumn{3}{c}{Main drift chamber (MDC)} \\
		\hline
		42 & \verb|LTrk_BB|    & \phantom{1}1.2\% \\
		43 & \verb|LTrk.GE.N|  & \phantom{1}0.08\% \\
		44 & \verb|LTrk.GE.2|  & \phantom{1}3.8\% \\
		45 & \verb|LTrk.GE.1|  & 12.0\% \\	
		\bottomrule
\end{tabular*}
\end{center}


\vspace{-1mm}
\centerline{\rule{80mm}{0.1pt}}
\vspace{2mm}



\end{multicols}

\clearpage

\end{document}